\documentclass[10pt,a4paper,twocolumn,english,nofootinbib]{revtex4}
\usepackage{lmodern}
\usepackage{lmodern}

\usepackage[T1]{fontenc}
\usepackage[latin9]{inputenc}
\usepackage{babel}
\usepackage{amsmath}
\usepackage{amssymb}
\usepackage{esint}
\usepackage[unicode=true,pdfusetitle,
 bookmarks=true,bookmarksnumbered=false,bookmarksopen=false,
 breaklinks=false,pdfborder={0 0 1},backref=false,colorlinks=false]
 {hyperref}

\makeatletter

\@ifundefined{textcolor}{}
{%
 \definecolor{BLACK}{gray}{0}
 \definecolor{WHITE}{gray}{1}
 \definecolor{RED}{rgb}{1,0,0}
 \definecolor{GREEN}{rgb}{0,1,0}
 \definecolor{BLUE}{rgb}{0,0,1}
 \definecolor{CYAN}{cmyk}{1,0,0,0}
 \definecolor{MAGENTA}{cmyk}{0,1,0,0}
 \definecolor{YELLOW}{cmyk}{0,0,1,0}
}

\usepackage{babel}
\usepackage{graphicx}
\def\b{\begin{equation}}
\def\e{\end{equation}}

\@ifundefined{textcolor}{}{%
 \definecolor{BLACK}{gray}{0}
 \definecolor{WHITE}{gray}{1}
 \definecolor{RED}{rgb}{1,0,0}
 \definecolor{GREEN}{rgb}{0,1,0}
 \definecolor{BLUE}{rgb}{0,0,1}
 \definecolor{CYAN}{cmyk}{1,0,0,0}
 \definecolor{MAGENTA}{cmyk}{0,1,0,0}
 \definecolor{YELLOW}{cmyk}{0,0,1,0}
 }

\usepackage{latexsym}\usepackage{bm}

\makeatother

\begin{document}
\title{Conserved charges in $AdS$: A new formula}
\author{Emel Altas}
\email{emelaltas@kmu.edu.tr}

\affiliation{Department of Physics,\\
 Karamanoglu Mehmetbey University, 70100, Karaman, Turkey}
\author{Bayram Tekin}
\email{btekin@metu.edu.tr}

\affiliation{Department of Physics,\\
 Middle East Technical University, 06800, Ankara, Turkey}
\date{\today}
\begin{abstract}
We give a new construction of conserved charges in asymptotically anti-de
Sitter spacetimes in Einstein's gravity. The new formula is explicitly gauge-invariant and makes direct use of the linearized curvature tensor instead of the metric perturbation. As an example, we compute the mass and angular momentum of the Kerr-AdS
black holes. 
\end{abstract}
\maketitle

\section{Introduction}

In Einstein's gravity, outside a source, in a vacuum, all the effects of gravity are encoded
in the Riemann tensor (or the Weyl tensor when there is no cosmological
constant). This should also be the case for conserved charges, such
as mass-energy and angular momentum. Here we show that such a construction
of conserved charges exists in asymptotically anti de Sitter $(AdS)$
spacetimes. Namely the total mass-energy or angular momentum of an
asymptotically $AdS$ spacetime can be directly computed from an integral
that is written in terms of the linearized part of the Riemann tensor.

Just like in any other theory in a flat spacetime, conserved charges
in gravity play a major role in understanding the integration parameters
that appear in the classical solutions such as the black holes and their
thermodynamics. But in contrast to the flat spacetime, a generic
curved spacetime does not have any symmetries and hence one should
not expect any conserved quantities. Fortunately, for some spacetimes
which are important in black hole physics and cosmology, one can define
total mass (energy) and angular momentum given that the spacetime
is asymptotically flat or (anti)-de-Sitter. For an asymptotically flat
spacetime, we have the celebrated Arnowitt-Deser-Misner (ADM) mass
\citep{adm} which is also a geometric invariant for the spacelike hypersurface
of the four dimensional spacetime as long as certain asymptotic conditions
on the decay of the metric tensor and the extrinsic curvature are
satisfied. One can also give a similar formula for the total angular
momentum of asymptotically flat spacetimes. A generalization to asymptotically
$(A)dS$ spacetimes was carried out by Abbott and Deser (AD) \citep{Abbott_Deser}.

In the usual formulation of conserved charges \cite{Deser_Tekin}, given a background Killing vector $\bar{\xi}^{\mu}$ a partially conserved current in Einstein's theory can be found as 
\begin{equation}
J^{\mu}:=\sqrt{-g}\bar{\text{\ensuremath{\xi}}}_{\nu}\left(\text{\ensuremath{{\cal {G}}}}^{\nu\mu}\right)^{\left(1\right)}, \hskip 0.5 cm \partial_{\mu}J^{\mu}=0,
\end{equation}
where $\left(\text{\ensuremath{{\cal {G}}}}^{\nu\mu}\right)^{\left(1\right)}$ is the linearized cosmological Einstein tensor and the linearization of the field equations read
 $\left(\text{\ensuremath{{\cal {G}}}}^{\nu\mu}\right)^{\left(1\right)}=\kappa\tau^{\mu\nu}+{\cal {O}}(h^{2},h^{3},...) =: \kappa T^{\mu\nu}$.
So the conserved charge is  
\begin{equation}
Q(\ensuremath{\xi}):=\intop_{\bar \Sigma}d^{n-1}y \thinspace\sqrt{-g}\thinspace\bar{\text{\ensuremath{\xi}}}_{\nu}\left(\text{\ensuremath{{\cal {G}}}}^{\nu0}\right)^{\left(1\right)},
\label{ilk}
\end{equation}
where $\bar \Sigma$ is a spatial hypersurface.  Note that as $T^{\mu \nu}$ includes all the localized matter and higher order gravitational corrections, despite appearance, (\ref{ilk}) captures all the non-linear terms.  See the recent review articles \cite{Adami, Compere}
for more details. To proceed further one
needs to write $\bar{\text{\ensuremath{\xi}}}_{\nu}\left(\text{\ensuremath{{\cal {G}}}}^{\nu\mu}\right)^{\left(1\right)}$
to be the divergence of a tensor. This requires writing $\left(\text{\ensuremath{{\cal {G}}}}^{\nu\mu}\right)^{\left(1\right)}$
explicitly in terms of the metric perturbation $h_{\mu\nu}$ which yields \cite{Abbott_Deser}
\begin{equation}
\bar{\text{\ensuremath{\xi}}}_{\nu}\left(\text{\ensuremath{{\cal {G}}}}^{\nu\mu}\right)^{\left(1\right)}=\nabla_{\alpha}\left(\bar{\text{\ensuremath{\xi}}}_{\nu}\nabla_{\beta}K^{\mu\alpha\nu\beta}-K^{\mu\beta\nu\alpha}\nabla_{\beta}\bar{\text{\ensuremath{\xi}}}_{\nu}\right),
\label{iki}
\end{equation}
with the superpotential given as
\begin{equation}
K^{\mu\alpha\nu\beta}:=\frac{1}{2}\left(\bar{g}^{\alpha\nu}\tilde{h}^{\mu\beta}+\bar{g}^{\mu\beta}\tilde{h}^{\alpha\nu}-\bar{g}^{\alpha\beta}\tilde{h}^{\mu\nu}-\bar{g}^{\mu\nu}\tilde{h}^{\alpha\beta}\right),
\end{equation}
and $\tilde{h}^{\mu\nu}:=h^{\mu\nu}-\frac{1}{2}\bar{g}^{\mu\nu}h.$ The crux of the above construction is that one must use the explicit form the linearized Einstein tensor in terms of the metric perturbation (or deviation from the $(A)dS$ background).  This yields (\ref{iki}) which is invariant under gauge transformations of the form  $\delta h_{\mu \nu} = \bar \nabla_\mu \zeta_\nu +\bar \nabla_\nu \zeta_\mu$, but neither $K^{\mu\alpha\nu\beta}$ , nor the two-from current 
$\left(\bar{\text{\ensuremath{\xi}}}_{\nu}\nabla_{\beta}K^{\mu\alpha\nu\beta}-K^{\mu\beta\nu\alpha}\nabla_{\beta}\bar{\text{\ensuremath{\xi}}}_{\nu}\right)$ that appears in the right-hand side of (\ref{iki}) are gauge invariant: under these transformations a boundary terms appears \cite{Emel_PRD_uzun}. So, even though the charge $Q(\ensuremath{\xi})$ is gauge invariant, the integrand defining the charge is not. The question is if one can find a way to make all this construction explicitly gauge invariant. The answer is not obvious because not every gauge-invariant physical quantity can be written explicitly gauge-invariant in local way. 

To achieve our goal of finding a fully gauge-invariant expression, here we shall provide another method of expressing $\bar{\text{\ensuremath{\xi}}}_{\nu}\left(\text{\ensuremath{{\cal {G}}}}^{\nu\mu}\right)^{\left(1\right)}$ in
such a way that one does not explicitly use the expression of $\left(\text{\ensuremath{{\cal {G}}}}^{\nu\mu}\right)^{\left(1\right)}$, instead purely geometric considerations
will be used such that the charges are expressed in terms of the linearized
Riemann tensor. The formula, whose derivation will be given below, reads as 
\begin{equation}
Q\left(\bar{\xi}\right)=k\int_{\partial\bar{\Sigma}}d^{n-2}x\,\sqrt{\bar{\gamma}}\,\bar{\epsilon}_{\mu\nu}\left(R^{\nu\mu}\thinspace_{\beta\sigma}\right)^{\left(1\right)}\bar{\text{\ensuremath{{\cal {F}}}}}^{\beta\sigma},
\label{newcharge}
\end{equation}
with the constant coefficient 
\begin{equation}
k=\frac{(n-1)(n-2)}{8(n-3)\Lambda G\Omega_{n-2}}
\label{katsayi}
\end{equation}
and $(R^{\nu\mu}\thinspace_{\beta\sigma})^{\left(1\right)}$ is the
linearized part of the Riemann tensor about the $AdS$ background. All
the barred quantities refer to the background spacetime $\bar{{\cal {M}}}$
whose boundary is $\partial\bar{{\cal {M}}}$. The Killing vector
is $\bar{\xi}^{\sigma}$ and the antisymmetric tensor is $\bar{{\cal {F}}}^{\beta\sigma}:=\bar{\nabla}^{\beta}\bar{\xi}^{\sigma}$.
$\bar{\Sigma}$ is a spatial hypersurface which is not equal to $\partial\bar{{\cal {M}}}$,
hence $\bar{\Sigma}$ can have a boundary of its own which is $\partial\bar{\Sigma}$.
Here 
\begin{equation}
\bar{\epsilon}_{\mu\nu}:=\frac{1}{2}\left(\bar{n}_{\mu}\bar{\sigma}_{\nu}-\bar{n}_{\nu}\bar{\sigma}_{\mu}\right),
\end{equation}
where $\bar{n}_{\mu}$ is a normal one form on $\partial\bar{{\cal {M}}}$
and $\bar{\sigma}_{\nu}$ is the unit normal one form on $\partial\bar{\Sigma}$
and $\bar{\gamma}$ is the induced metric  on the boundary.

\section{Derivation of the new formula}

Let us now provide the derivation of (\ref{newcharge}): we start with the second Bianchi identity 
\begin{equation}
\nabla_{\nu}R_{\sigma\beta\mu\rho}+\nabla_{\sigma}R_{\beta\nu\mu\rho}+\nabla_{\beta}R_{\nu\sigma\mu\rho}=0,
\end{equation}
multiplying with $g^{\nu\rho}$ and making use of the definition of
the cosmological Einstein tensor ${\cal {G}}_{\beta}^{\nu}:=R_{\beta}^{\nu}-\frac{1}{2}R\delta_{\nu}^{\beta}+\Lambda\delta_{\nu}^{\beta}$,
one arrives at 
\begin{equation}
\nabla_{\nu}{\cal{P}}^{\nu}\thinspace_{\mu\beta\sigma}=0,\label{divK}
\end{equation}
where the ${\cal{P}}$ tensor reads 
\begin{multline}
\text{\ensuremath{{\cal{P}}}}^{\nu}\thinspace_{\mu\beta\sigma}:=R^{\nu}\thinspace_{\mu\beta\sigma}+\delta_{\sigma}^{\nu}\text{\ensuremath{{\cal {G}}}}_{\beta\mu}-\delta_{\beta}^{\nu}\text{\ensuremath{{\cal {G}}}}_{\sigma\mu}+\text{\ensuremath{{\cal {G}}}}_{\sigma}^{\nu}g_{\beta\mu}-\text{\ensuremath{{\cal {G}}}}_{\beta}^{\nu}g_{\sigma\mu}\\
+(\frac{R}{2}-\frac{\Lambda(n+1)}{n-1})(\delta_{\sigma}^{\nu}g_{\beta\mu}-\delta_{\beta}^{\nu}g_{\sigma\mu}).
\end{multline}
In the construction of this tensor we have used  $\nabla_{\mu}{\cal {G}}^{\mu\nu}=0$
and $\nabla_{\mu}g^{\mu\nu}=0$ and defined it in such a way that its $AdS$ value vanishes. For any smooth metric, (\ref{divK})
is valid {\it identically} without the use of the field equations. We can
also express the ${\cal{P}}$ tensor in terms of the Weyl tensor
as 
\begin{multline}
\text{\ensuremath{{\cal{P}}}}^{\nu\mu\beta\sigma}=C^{\nu\mu\beta\sigma}-\frac{2(n-3)}{n-2}(\text{\ensuremath{{\cal {G}}}}^{\nu[\beta}g^{\sigma]\mu}+\text{\ensuremath{{\cal {G}}}}^{\mu[\sigma}g^{\beta]\nu})\\
+\frac{n-3}{n-1}(\frac{\Lambda n}{n-2}-\frac{R}{2})(g^{\nu\beta}g^{\mu\sigma}-g^{\mu\beta}g^{\nu\sigma}).
\end{multline}
Let $\text{\ensuremath{{\cal {F}}}}^{\beta\sigma}$ be a generic antisymmetric
tensor. Then, contracting (\ref{divK}) with $\text{\ensuremath{{\cal {F}}}}_{\beta\sigma}$
yields 
\begin{equation}
\nabla_{\nu}(\text{\ensuremath{{\cal {F}}}}_{\beta\sigma}\text{\ensuremath{{\cal{P}}}}^{\nu\mu\beta\sigma})-\text{\ensuremath{{\cal{P}}}}^{\nu\mu\beta\sigma}\nabla_{\nu}\text{\ensuremath{{\cal {F}}}}_{\beta\sigma}=0,\label{eq:ddimensionalmainequation}
\end{equation}
which is an exact equation. Let us now consider the metric perturbation which defines asymptotically $AdS$ spacetimes
\begin{equation}
g_{\mu\nu}=\overline{g}_{\mu\nu}+ h_{\mu\nu},
\end{equation}
where the background metric is $AdS$ and satisfies 
\begin{equation}
\bar{R}_{\alpha\beta\gamma\delta}=\frac{2\Lambda}{(n-1)(n-2)}\left(\overline{g}_{\alpha\gamma}\overline{g}_{\beta\delta}-\overline{g}_{\alpha\delta}\overline{g}_{\beta\gamma}\right),
\end{equation}
together with Ricci tensor $\bar{R}_{\alpha\beta}=\frac{2\Lambda}{n-2}\overline{g}_{\alpha\beta}$
and the scalar curvature $\bar{R}=\frac{2\Lambda n}{n-2}$. For the
AdS background we have $\bar{\text{\ensuremath{{\cal {G}}}}}^{\mu\nu}$= 0
and $\bar{\text{\ensuremath{{\cal{P}}}}}^{\nu\mu\beta\sigma}=0$ as already noted.
Let us now consider the following particular anti-symmetric tensor 
\begin{equation}
\text{\ensuremath{{\cal {F}}}}_{\alpha\beta}:=\frac{1}{2}\left(\nabla_{\alpha}\xi_{\beta}-\nabla_{\beta}\text{\ensuremath{\xi}}_{\alpha}\right).
\end{equation}
When $\text{\ensuremath{\xi}}$ is a background Killing vector one
has $\text{\ensuremath{{\cal {F}}}}_{\alpha\beta}=\bar{\text{\ensuremath{{\cal {F}}}}}_{\alpha\beta}$.
The linear order expansion of (\ref{eq:ddimensionalmainequation})
reads 
\begin{equation}
\bar{\nabla_{\nu}}((\text{\ensuremath{{\cal{P}}}}^{\nu\mu\beta\sigma})^{(1)}\bar{\text{\ensuremath{{\cal {F}}}}}_{\beta\sigma})-(\text{\ensuremath{{\cal{P}}}}^{\nu\mu\beta\sigma})^{(1)}\bar{\nabla}_{\nu}\bar{\text{\ensuremath{{\cal {F}}}}}_{\beta\sigma}=0.\label{eq:ddimensionalmainequationlinear}
\end{equation}
We now need to calculate the first order linearization of the ${\cal{P}}$ tensor which reads
\begin{multline}
({\cal{P}}^{\nu\mu\beta\sigma})^{\left(1\right)}=(R^{\nu\mu\beta\sigma})^{1}+2(\text{\ensuremath{{\cal {G}}}}^{\mu[\beta})^{(1)}\overline{g}^{\sigma]\nu}+2 (\text{\ensuremath{{\cal {G}}}}^{\nu[\sigma})^{(1)}\overline{g}^{\beta]\mu}\\
+(R)^{\left(1\right)}\overline{g}^{\mu[\beta}\overline{g}^{\sigma]\nu}
+\frac{4\Lambda}{(n-1)(n-2)}(h^{\mu[\sigma}\overline{g}^{\beta]\nu}+
\overline{g}^{\mu[\sigma}{h}^{\beta]\nu}).\label{ktensorlinear}
\end{multline}
After some manipulations and using the identity $\bar{\nabla}_{\mu}\bar{\nabla}_{\nu}\text{\ensuremath{\bar{\xi}}}_{\rho}=\bar{R}^{\sigma}\thinspace_{\mu\nu\rho}\,\text{\ensuremath{\bar{\xi}}}_{\sigma}$, 
one arrives at 
\begin{equation}
(\text{\ensuremath{{\cal{P}}}}^{\nu\mu\beta\sigma})^{(1)}\bar{\nabla}_{\nu}\bar{\text{\ensuremath{{\cal {F}}}}}_{\beta\sigma}=\frac{4\Lambda (n-3)}{(n-1)(n-2)}\bar{\text{\ensuremath{\xi}}}_{\lambda}(\text{\ensuremath{{\cal {G}}}}^{\lambda\mu})^{(1)},
\end{equation}
then from (\ref{eq:ddimensionalmainequationlinear}) we obtain the
main expression 
\begin{equation}
\bar{\text{\ensuremath{\xi}}}_{\lambda}(\text{\ensuremath{{\cal {G}}}}^{\lambda\mu})^{(1)}=\frac{(n-1)(n-2)}{4\Lambda (n-3)}\bar{\nabla_{\nu}}\Big ((\text{\ensuremath{{\cal{P}}}}^{\nu\mu\beta\sigma})^{(1)}\bar{\text{\ensuremath{{\cal {F}}}}}_{\beta\sigma}\Big).\label{eq:finallinearequation}
\end{equation}
This proves our desired formula: without writing the linearized Einstein tensor explicitly in terms of the metric perturbation, we were able to express the conserved current as a boundary term involving the linearization of the Riemann and Einstein tensors as well as the Ricci scalar.  
To arrive at the total charge expression, we use the Stokes' theorem and the resulting integral must be evaluated at spatial infinity. This simplifies the expression further: $\left(\text{\ensuremath{{\cal {G}}}}^{\nu\mu}\right)^{\left(1\right)}$ and the linearized scalar curvature vanishes at infinity. Moreover lowering the last two indices of the  $(\text{\ensuremath{{\cal{P}}}}^{\nu\mu\beta\sigma})^{(1)}$ tensor one arrives at the charge expression (\ref{newcharge}).

\subsection*{Application to Kerr-AdS black holes}

As an application of our formula, let us consider the  Kerr-$AdS$ black hole in four dimensions. 
One can take the solution to be in the Kerr-Schild form which reads
\begin{equation}
ds^{2}=d\bar{s}^{2}+\frac{2M r }{\rho^2}\left(k_{\mu}dx^{\mu}\right)^{2},
\end{equation}
where $
\rho^{2}=r^{2}+a^{2}\cos^{2}\theta$ and 
with the $AdS$ seed metric given as  
\begin{eqnarray}
d\bar{s}^{2} & = & -\frac{\left(1-\frac{\Lambda r^{2}}{3}\right)\Delta_{\theta}dt^{2}}{\left(1+\frac{\Lambda a^{2}}{3}\right)}+\frac{\rho^{2}dr^{2}}{\left(1-\frac{\Lambda r^{2}}{3}\right)\left(r^{2}+a^{2}\right)}\nonumber \\
 & + & \frac{\rho^{2}d\theta^{2}}{\Delta_{\theta}}+\frac{\left(r^{2}+a^{2}\right)\sin^{2}\theta d\phi^{2}}{\left(1+\frac{\Lambda a^{2}}{3}\right)},
\end{eqnarray}
where $\Delta_{\theta} = 1+ \frac{\Lambda}{3} \cos^2 \theta$. 
The null vector $k_{\mu}$ is given by
\begin{equation}
k_{\mu}dx^{\mu}=\frac{\Delta_{\theta}dt}{\left(1+\frac{\Lambda a^{2}}{3}\right)}+\frac{\rho^{2}dr}{\left(1-\frac{\Lambda r^{2}}{3}\right)\left(r^{2}+a^{2}\right)}-\frac{a \sin^{2}\theta d\phi}{\left(1+\frac{\Lambda a^{2}}{3}\right)}.
\nonumber 
\end{equation}
Taking the Killing vector to be $ \bar \xi = ( -1,0,0,0)$, and $G=1$, the charge expression  (\ref{newcharge})
becomes 
\begin{equation}
E = \frac{3}{ 16 \pi \Lambda} \int_{S^2_\infty} d\Omega  (R^{r t}\thinspace_{\beta\sigma})^{\left(1\right)}\bar{\nabla}^\beta\bar{\xi}^\sigma,
\end{equation}
with $\sqrt{\bar \gamma} = \frac{r^2+ a^2 \cos^2\theta}{1+ \frac{\Lambda}{3} a^2}$. The integral is over a sphere at $r \rightarrow \infty$ which yields the answer
\begin{equation}
E=\frac{M}{\left(1+\frac{\Lambda a^{2}}{3}\right)^{2}}.
\end{equation}
Similarly for the Killing vector $\bar \xi = ( 0,0,0,1)$ one finds the angular momentum of the black hole as
\begin{equation}
J=\frac{aM}{\left(1+\frac{\Lambda a^{2}}{3}\right)^{2}}.
\end{equation}
These relations satisfy $E=J/a$ and they match the ones computed in \cite{Kanik}.

\section{Relation of the new formula with the Abbott-Deser formula}

Let us derive the explicit connection between the AD expression (\ref{iki}) and the one we have given here (\ref{eq:finallinearequation}). Going from the former to the latter is extremely difficult, one needs judicious additions of terms that vanish, so we shall start from our expression and expand it to find out the relation.  For this purpose, let us start from the linearized form of the $(2,2)$ background tensor 
\begin{multline}
\text{\ensuremath{{\cal {P}}}}^{\nu\mu}\thinspace_{\beta\sigma}:=R^{\nu\mu}\thinspace_{\beta\sigma}+\delta_{\sigma}^{\nu}(R_{\beta}^{\mu})^{\left(1\right)}-\delta_{\beta}^{\nu}(R_{\sigma}^{\mu})^{\left(1\right)}+\delta_{\beta}^{\mu}(R_{\sigma}^{\nu})^{\left(1\right)}\\
-\delta_{\sigma}^{\mu}(R_{\beta}^{\nu})^{\left(1\right)}-\frac{1}{2}(R)^{\left(1\right)}(\delta_{\sigma}^{\nu}\delta_{\beta}^{\mu}-\delta_{\beta}^{\nu}\delta_{\sigma}^{\mu}),\label{Ptensorlinear}
\end{multline}
which, due to the symmetries, yields 
\begin{multline}
(\text{\ensuremath{{\cal {P}}}}^{\nu\mu}\thinspace_{\beta\sigma})^{(1)}\bar{\text{\ensuremath{{\cal {F}}}}}^{\beta\sigma}=\bar{\text{\ensuremath{{\cal {F}}}}}^{\beta\sigma}(R^{\nu\mu}\thinspace_{\beta\sigma})^{(1)}\\
+2\bar{\text{\ensuremath{{\cal {F}}}}}^{\sigma\nu}(R_{\sigma}^{\mu})^{(1)}-2\bar{\text{\ensuremath{{\cal {F}}}}}^{\sigma\mu}(R_{\sigma}^{\nu})^{(1)}-\bar{\text{\ensuremath{{\cal {F}}}}}^{\mu\nu}(R)^{(1)}.\label{eq:newformulationtoad}
\end{multline}
Let us compute the right-hand side of the last expression term by
term. The first term can be written as
\begin{multline}
\bar{\text{\ensuremath{{\cal {F}}}}}^{\beta\sigma}(R^{\nu\mu}\thinspace_{\beta\sigma})^{(1)}=\frac{1}{2}\bar{\text{\ensuremath{{\cal {F}}}}}^{\beta\sigma}\left (-\bar{R}^{\nu}\thinspace_{\lambda\beta\sigma}h^{\lambda\mu}+\bar{R}^{\mu}\thinspace_{\lambda\beta\sigma}h^{\lambda\nu}\right.\\
\left.+\bar{g}^{\lambda\mu}(R^{\nu}\thinspace_{\lambda\beta\sigma})^{(1)}-\bar{g}^{\lambda\nu}(R^{\mu}\thinspace_{\lambda\beta\sigma})^{(1)}\right).
\end{multline}
Using the first order linearized Riemann tensor
\begin{equation}
(R^{\nu}\thinspace_{\lambda\beta\sigma})^{(1)}=\bar{\nabla}_{\beta}(\Gamma_{\lambda\sigma}^{\nu})^{\left(1\right)}-\bar{\nabla}_{\sigma}(\Gamma_{\lambda\beta}^{\nu})^{\left(1\right)},
\end{equation}one finds 
\begin{multline}
\bar{\text{\ensuremath{{\cal {F}}}}}^{\beta\sigma}(R^{\nu\mu}\thinspace_{\beta\sigma})^{(1)}=\frac{2\Lambda}{(n-1)(n-2)}(\bar{\text{\ensuremath{{\cal {F}}}}}^{\mu\sigma}h_{\sigma}^{\nu}-\bar{\text{\ensuremath{{\cal {F}}}}}^{\nu\sigma}h_{\sigma}^{\mu})\\
+\bar{\text{\ensuremath{{\cal {F}}}}}^{\beta\sigma}\bar{\nabla}_{\beta}(\bar{\nabla}^{\mu}h_{\sigma}^{\nu}-\bar{\nabla}^{\nu}h_{\sigma}^{\mu}).
\end{multline}
We can rewrite this as follows:
\begin{multline}
\bar{\text{\ensuremath{{\cal {F}}}}}^{\beta\sigma}(R^{\nu\mu}\thinspace_{\beta\sigma})^{(1)}=\frac{2\Lambda}{(n-1)(n-2)}\left(h_{\sigma}^{\nu}\bar{\nabla}^{\mu}\bar{\text{\ensuremath{\xi}}}^{\sigma}\right.\\
\left.-h_{\sigma}^{\mu}\bar{\nabla}^{\nu}\bar{\text{\ensuremath{\xi}}}^{\sigma}+(n-1)\bar{\text{\ensuremath{\xi}}}^{\sigma}\bar{\nabla}^{\mu}h_{\sigma}^{\nu}-(n-1)\bar{\text{\ensuremath{\xi}}}^{\sigma}\bar{\nabla}^{\nu}h_{\sigma}^{\mu}\right)\\
+\bar{\nabla}_{\beta}\left(\bar{\text{\ensuremath{{\cal {F}}}}}^{\beta\sigma}(\bar{\nabla}^{\mu}h_{\sigma}^{\nu}-\bar{\nabla}^{\nu}h_{\sigma}^{\mu})\right)
\end{multline}
Now, we can compute the second term in (\ref{eq:newformulationtoad}) as
\begin{equation}
2\bar{\text{\ensuremath{{\cal {F}}}}}^{\sigma\nu}(R_{\sigma}^{\mu})^{(1)}=2\bar{\text{\ensuremath{{\cal {F}}}}}^{\sigma\nu}\Big (\bar{g}^{\lambda\mu}(R_{\lambda\sigma})^{\left(1\right)}-h^{\lambda\mu}\bar{R}_{\lambda\sigma})\Big)
\end{equation}
where
\begin{equation}
(R_{\lambda\sigma})^{\left(1\right)}=\frac{1}{2}(\bar{\nabla}_{\rho}\bar{\nabla}_{\lambda}h_{\sigma}^{\rho}+\bar{\nabla}_{\rho}\bar{\nabla}_{\sigma}h_{\lambda}^{\rho}-\bar{\nabla}_{\rho}\bar{\nabla}^{\rho}h_{\lambda\sigma}-\bar{\nabla}_{\lambda}\bar{\nabla}_{\sigma}h)
\end{equation}
Then we have 
\begin{multline}
2\bar{\text{\ensuremath{{\cal {F}}}}}^{\sigma\nu}(R_{\sigma}^{\mu})^{(1)}=\frac{2\Lambda}{(n-1)(n-2)}\Bigl(\bar{\text{\ensuremath{\xi}}}^{\sigma}\bar{\nabla}^{\nu}h_{\sigma}^{\mu}-h\bar{\nabla}^{\mu}\bar{\text{\ensuremath{\xi}}}^{\nu}\\
-\bar{\text{\ensuremath{\xi}}}^{\sigma}\bar{\nabla}^{\mu}h_{\sigma}^{\nu}+(n-2)(h_{\sigma}^{\mu}\bar{\nabla}^{\nu}\bar{\text{\ensuremath{\xi}}}^{\sigma}+\bar{\text{\ensuremath{\xi}}}^{\nu}\bar{\nabla}_{\sigma}h^{\sigma\mu}-\bar{\text{\ensuremath{\xi}}}^{\nu}\bar{\nabla}^{\mu}h)\Bigr) \\
+\bar{\nabla}_{\rho}\left(\bar{\text{\ensuremath{{\cal {F}}}}}^{\sigma\nu}(\bar{\nabla}^{\mu}h_{\sigma}^{\rho}+\delta_{\sigma}^{\rho}\bar{\nabla}_{\beta}h^{\beta\mu}-\bar{\nabla}^{\rho}h_{\sigma}^{\mu}-\delta_{\sigma}^{\rho}\bar{\nabla}^{\mu}h)\right).
\end{multline}
Finally we can compute the last term in (\ref{eq:newformulationtoad}) as
\begin{multline}
\bar{\text{\ensuremath{{\cal {F}}}}}^{\mu\nu}(R)^{(1)}=\frac{2\Lambda}{(n-1)(n-2)}\Bigl(-\bar{\text{\ensuremath{\xi}}}^{\mu}\bar{\nabla}_{\sigma}h^{\sigma\nu}+\bar{\text{\ensuremath{\xi}}}^{\mu}\bar{\nabla}^{\nu}h\\
+\bar{\text{\ensuremath{\xi}}}^{\nu}\bar{\nabla}_{\sigma}h^{\sigma\mu}-\bar{\text{\ensuremath{\xi}}}^{\nu}\bar{\nabla}^{\mu}h-(n-1)h\bar{\nabla}^{\mu}\bar{\text{\ensuremath{\xi}}}^{\nu}\Bigr) \\
\bar{\nabla}_{\rho}\left(\bar{\text{\ensuremath{{\cal {F}}}}}^{\mu\nu}(\bar{\nabla}_{\sigma}h^{\rho\sigma}-\bar{\nabla}^{\rho}h)\right)\\
\end{multline}
Collecting all the pieces together, we have the following expression 
\begin{multline}
(\text{\ensuremath{{\cal {P}}}}^{\nu\mu}\thinspace_{\beta\sigma})^{(1)}\bar{\text{\ensuremath{{\cal {F}}}}}^{\beta\sigma}=\frac{4\Lambda(n-3)}{(n-1)(n-2)}\Bigl(h_{\sigma}^{[\mu}\bar{\nabla}^{\nu]}\bar{\text{\ensuremath{\xi}}}^{\sigma}+\bar{\text{\ensuremath{\xi}}}^{\sigma}\bar{\nabla}^{[\mu}h_{\sigma}^{\nu]}\\
+\bar{\text{\ensuremath{\xi}}}^{[\nu}\bar{\nabla}_{\sigma}h^{\mu] \sigma}+\bar{\text{\ensuremath{\xi}}}^{[\mu}\bar{\nabla}^{\nu]}h+\frac{1}{2}h\bar{\nabla}^{\mu}\bar{\text{\ensuremath{\xi}}}^{\nu}\Bigr) \\
+\bar{\nabla}_{\rho}\left(\bar{\text{\ensuremath{{\cal {F}}}}}^{\sigma\nu}(\bar{\nabla}^{\mu}h_{\sigma}^{\rho}+\delta_{\sigma}^{\rho}\bar{\nabla}_{\beta}h^{\beta\mu}-\bar{\nabla}^{\rho}h_{\sigma}^{\mu}-\delta_{\sigma}^{\rho}\bar{\nabla}^{\mu}h)\right.\\
\left. -\frac{1}{2}\bar{\text{\ensuremath{{\cal {F}}}}}^{\mu\nu}(\bar{\nabla}_{\sigma}h^{\rho\sigma}-\bar{\nabla}^{\rho}h)+ \bar{\text{\ensuremath{{\cal {F}}}}}^{\rho\sigma}\bar{\nabla}^{\mu}h_{\sigma}^{\nu} - (\mu \leftrightarrow \nu)
\right).
\end{multline}
from (\ref{eq:finallinearequation}), we can write
\begin{multline}
\bar{\text{\ensuremath{\xi}}}_{\lambda}(\text{\ensuremath{{\cal {G}}}}^{\lambda\mu})^{(1)}=\bar{\nabla_{\nu}}\Bigl(h_{\sigma}^{[\mu}\bar{\nabla}^{\nu]}\bar{\text{\ensuremath{\xi}}}^{\sigma}+\bar{\text{\ensuremath{\xi}}}^{\sigma}\bar{\nabla}^{[\mu}h_{\sigma}^{\nu]}\\
+\bar{\text{\ensuremath{\xi}}}^{[\nu}\bar{\nabla}_{\sigma}h^{\mu] \sigma}+\bar{\text{\ensuremath{\xi}}}^{[\mu}\bar{\nabla}^{\nu]}h+\frac{1}{2}h\bar{\nabla}^{\mu}\bar{\text{\ensuremath{\xi}}}^{\nu}\Bigr) \\
+\frac{(n-1)(n-2)}{4\Lambda (n-3)}\bar{\nabla_{\nu}}\bar{\nabla}_{\rho}\Bigl( -\frac{1}{2}\bar{\nabla}^{\mu}\bar{\text{\ensuremath{\xi}}}^{\nu}(\bar{\nabla}_{\sigma}h^{\rho\sigma}-\bar{\nabla}^{\rho}h)\\ 
+\bar{\nabla}^{\sigma}\bar{\text{\ensuremath{\xi}}}^{\nu} (\bar{\nabla}^{\mu}h_{\sigma}^{\rho}+\delta_{\sigma}^{\rho}\bar{\nabla}_{\beta}h^{\beta\mu}-\bar{\nabla}^{\rho}h_{\sigma}^{\mu} -\delta_{\sigma}^{\rho}\bar{\nabla}^{\mu}h)\\ +\bar{\nabla}^{\rho}\bar{\text{\ensuremath{\xi}}}^{\sigma}\bar{\nabla}^{\mu}h_{\sigma}^{\nu} 
 - (\mu \leftrightarrow \nu)\Bigr).
\label{lasteqn}
\end{multline}
The first two lines yield the AD expression as given  in \cite{Deser_Tekin} while the remaining part is of the form $\bar{\nabla}_{\nu}\bar{\nabla}_{\rho} Q^{\rho \mu \nu}[h]$. Integrating the above expression on a spatial hypersurface, after making use of the Stokes' theorem, the first two lines give the AD charge, while the other part having two derivatives remain a {\it total divergence} on the boundary of the hypersurface, vanishes since the boundary of the boundary is nil. Note that this equivalence does not work in 3 spacetime dimensions and for the asymptotically flat spacetimes. It is important to recognize the following: under gauge transformations, the left-hand side of (\ref{lasteqn}) is gauge invariant and so is the right-hand side. But, it is easy to see that the first two lines are gauge-invariant only up to a boundary term. 
Full gauge invariance is  recovered with the additional parts. The details of this discussion were given in \cite{Emel_PRD_uzun}. 

\section{Conclusions} 

We have given a conserved charge expression in Einstein's theory for asymptotically $(A)dS$ spacetimes which is directly written in terms of the linearized Riemann tensor and an anti-symmetric tensor that appears as the potential of the Killing vector on the boundary of the spatial hypersurface. The expression is explicitly gauge-invariant as the up-up-down-down linearized Riemann tensor is gauge invariant under small variations $\delta h_{\mu \nu} = \bar \nabla_\mu \zeta_\nu +\bar \nabla_\nu \zeta_\mu$. Our construction started from the second Bianchi Identity on the Riemann tensor and as such, the final expression of conserved charges is valid for $n >3$ and not valid for the case of three dimensions. A naive extension of  this construction to generic gravity theories as discussed in \cite{Deser_Tekin} is not that obvious and was carried out \cite{Emel_PRD_uzun} after this work appeared. Once a higher order theory's field equations are given one can work out a  similar computation for these theories and the coefficient $k$ in (\ref{katsayi}) receives corrections from the higher curvature terms. It would be interesting to relate our construction to the one given in \cite{kim}.

\end{document}